\documentclass{aastex}
\usepackage{spr-astr-addons}
\usepackage{url}

\RequirePackage{color}

\begin{document}

\title{Open Issues on the Synthesis of Evolved Stellar Populations at Ultraviolet Wavelengths}
\slugcomment{to appear in Astrophysics and Space Science Special volume}
\shorttitle{Stellar Populations in the UV}
\shortauthors{Chavez and Bertone}

\author{Miguel Chavez} \and \author{Emanuele Bertone}
\affil{Instituto Nacional de Astrofisica, Optica y Electronica, Luis
  E. Erro 1, 72840 Tonantzintla, Puebla, Mexico}
\email{mchavez@inaoep.mx}



\begin{abstract}
In this paper we briefly review three topics that have motivated our (and others') 
investigations in recent years within the context of evolutionary population 
synthesis techniques. These are: The origin of the FUV up-turn in elliptical galaxies,
the age-metallicity degeneracy, and the study of the mid-UV rest-frame spectra of distant 
red galaxies. We summarize some of our results and present a very preliminary application 
of a UV grid of theoretical spectra in the analysis of integrated properties of 
aged stellar populations. At the end, we concisely suggest how these topics can be tackled 
once the World Space Observatory enters into operation in the midst of this decade. 

\end{abstract}


\section{Introduction}
\label{sec:intro}

The study of stellar populations of non-resolved systems has greatly relied on
the models derived from evolutionary populations synthesis technique. This
approach is based on the spectrophometric properties of stars at, ideally, all
evolutionary phases and takes into account all phenomena that largely affect
the evolution of a star (e.g., mass-loss). Over the years, ever since the technique was 
first implemented \citep{tinsley68,tinsley72}, a wide variety of 
models based on different ingredients have been constructed and used in the
study of young and old stellar populations 
\citep[among the most popular ones,][]{buz89,bressan94,worthey94,bruzual93,bruzual03,leitherer99,maraston05}.
Not surprisingly, most of the work done up to date has vastly focused in the
optical spectrophotometric properties of stellar systems, and until 
relatively recently it has expanded to other wavelengths (as far as the
detailed analysis of spectral features is concerned), and, in some cases,
included the effects of an interstellar medium \citep{silva98,panuzzo05}.

At ultraviolet (UV) wavelengths, usually divided into two segments, the far-UV (1200--2000~\AA) 
and the mid-UV (2000--3200~\AA), the natural systems to look at are those whose underlying 
populations copiously emit and have their emission maxima in that window, i.e. star-forming 
systems. While these systems are extremely important in many astrophysical contexts 
\citep[see, e.g.,][]{buz02}, it was eventually realized that also old and intermediate age 
populations, which will be the main subject in this paper, deserve attention by their own right. 
As an example we can mention the countless studies motivated by the unexpected finding of a 
prominent far-UV flux excess in the bulge of Andromeda \citep{code69}. 
Aside of this far-UV flux excess, the mid-UV still remains vastly unexplored, in spite of the 
early suggestions that this wavelength region can help in lifting the so-called age-metallicity degeneracy (AMD) 
that plagues the optical spectrophotometric properties of evolved populations and that prevents 
the univocal determination of these parameters \citep{worthey94,dorman03}. Disentangling the effects of 
age and chemical composition is particularly important when attempting to evaluate the characteristics of 
distant red objects for which, through optical observations only feasible with the current generation of 
large telescopes, we can only access the rest-frame mid-UV flux \citep[e.g.,][]{dunlop96}.

Motivated by these three issues inherent to aged populations, the nature of the
far-UV flux, the AMD, and the properties of instrisically red galaxies up to
$z \sim 2$, we started a project aimed at providing complementary tools for
their analysis. In what follows, we present a short (and necessarily
incomplete) review of each of the above mentioned topics; we also briefly
describe our project and present some preliminary results that are still
under investigation.

\section{The Far-UV Rising Branch of Elliptical Galaxies}
\begin{figure}[!t]
\centering
\includegraphics[scale=0.80]{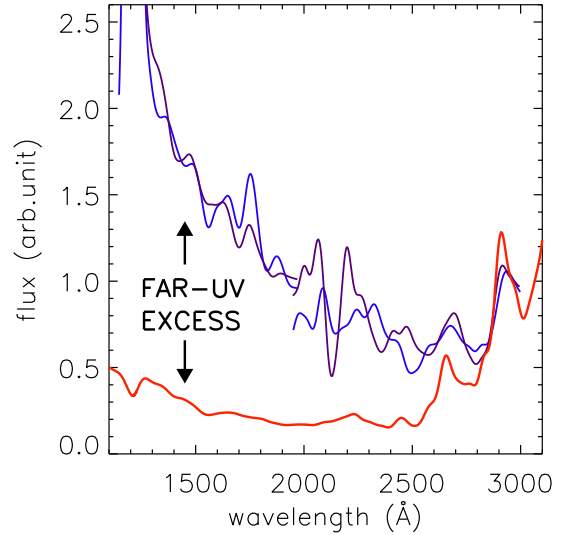}
\caption{Comparison of the mid-UV and far-UV spectral energy distribution of two elliptical galaxies 
observed with IUE (NGC1399 and NGC4649) to illustrate the far-UV excess. The
violet and blue curves correspond to the observed spectra, broadened with a Gaussian kernel of 50~\AA, and 
the red flux to a 10~Gyr old theoretical spectra of solar metallicity, calculated with the ingredients 
described in \citet{chavezetal09}.}
\label{fig:fuv_iue}
\end{figure}

Early-type stellar systems, such as elliptical galaxies and spiral bulges, are sometimes characterized
by a prominent flux shortward of 2000~\AA~(Fig.~\ref{fig:fuv_iue}). The nature of this flux excess (also called UV upturn, 
UVX phenomenon or UV rising branch) has been a subject of much debate ever since its discovery 
in the bulge of M31 \citep{code69}. Soon after this detection, a variety of hypotheses emerged
to explain the FUV rising flux \citep[see the excellent reviews by][]{burstein88,greggio90,oconnell99}: among them, a non-thermal 
origin through the galactic nuclear activity, the presence of an unexpected (for longly thought
quiescent systems) population of hot young stars \citep{tinsley72}, low-mass
metal-rich evolved objects \citep{bressan94,yi97}
or their metal-poor counterparts \citep{lee94}. More recently, it has been analyzed
the potential dominant role that subdwarf stars could play in modulating the
far-UV properties of the host galaxy, either as members of binary systems \citep{han07,han09}
or arising from the evolution of single stellar objects \citep{napi08,napi09}. 
A supplemental piece of the puzzle has been provided by the
detection of multiple main sequences (MSs) in galactic globular clusters, which reinforces the explanation of 
UVX phenomenon through presence of helium-rich sub-populations.

Over the past four decades evidence has grown in favour of the low-mass star
hypotheses. \citet{hills71} argued against a non-thermal origin based
on the overall shape of the far-UV energy distributions, which more closely
resembles that of the Rayleigh-Jeans tail of a thermal source. Additionally, the
diffuse distribution of the UV radiation, which can be fitted by a de
Vaucouleaurs profile such as the visible light, cannot be explained by the
presence of a highly concentrated source, as would be expected if an active
nucleus is the origin \citep{oke81,ohl98}. In a similar way, the UV imaginery 
has also worked against the residual star formation hypothesis (which would be implied by the
presence of hot MS stars) since, within the detection and resolution
limits of several experiments, O and B stars have not yet been detected.

Whilst the currently most accepted picture for the nature of the far-UV excess
is that the bulk of UV radiation is dominated  by helium burning low mass stars
and their progeny, in particular the so-called AGB-manqu\'e, the idea of having
on going star formation at very low levels has remained as a still plausible
elucidation for the UVX phenomenon \citep{rich05,rich09}. This fact might be supported by the
prevalence of molecular hydrogen (usually traced by the more accessible signatures of carbon monoxide
at millimeter wavelengths) 
in early-type systems, although in small amounts, indicative 
of star formation with an efficiency which is in fact similar to that found in spiral 
aggregates \citep[e.g.,][]{sage07}.

\section{The Age-Metallicity Degeneracy}

\begin{figure}[!t]
\centering
\begin{tabular}{c}
\includegraphics[scale=0.38]{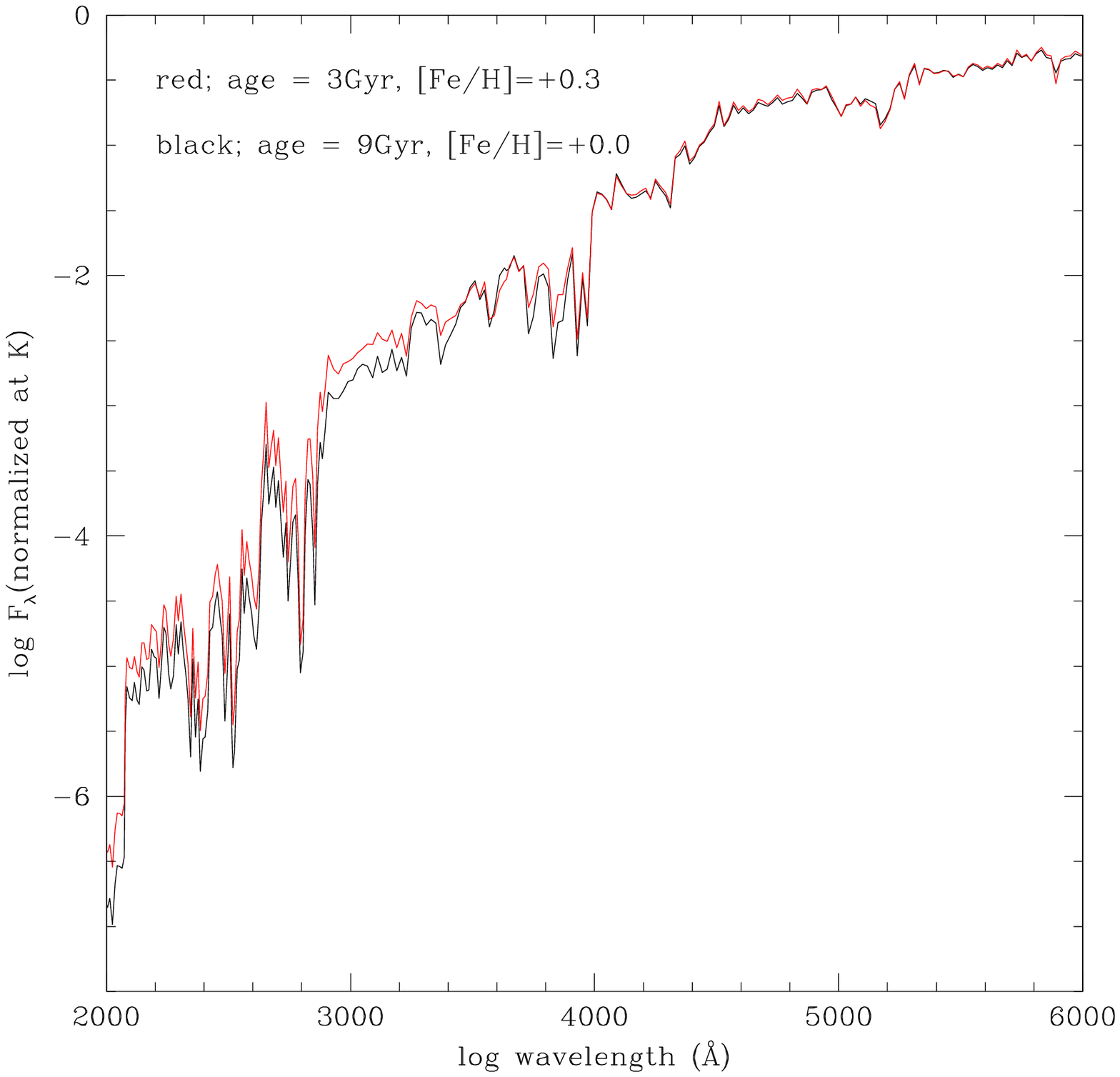} \\
\includegraphics[scale=0.38]{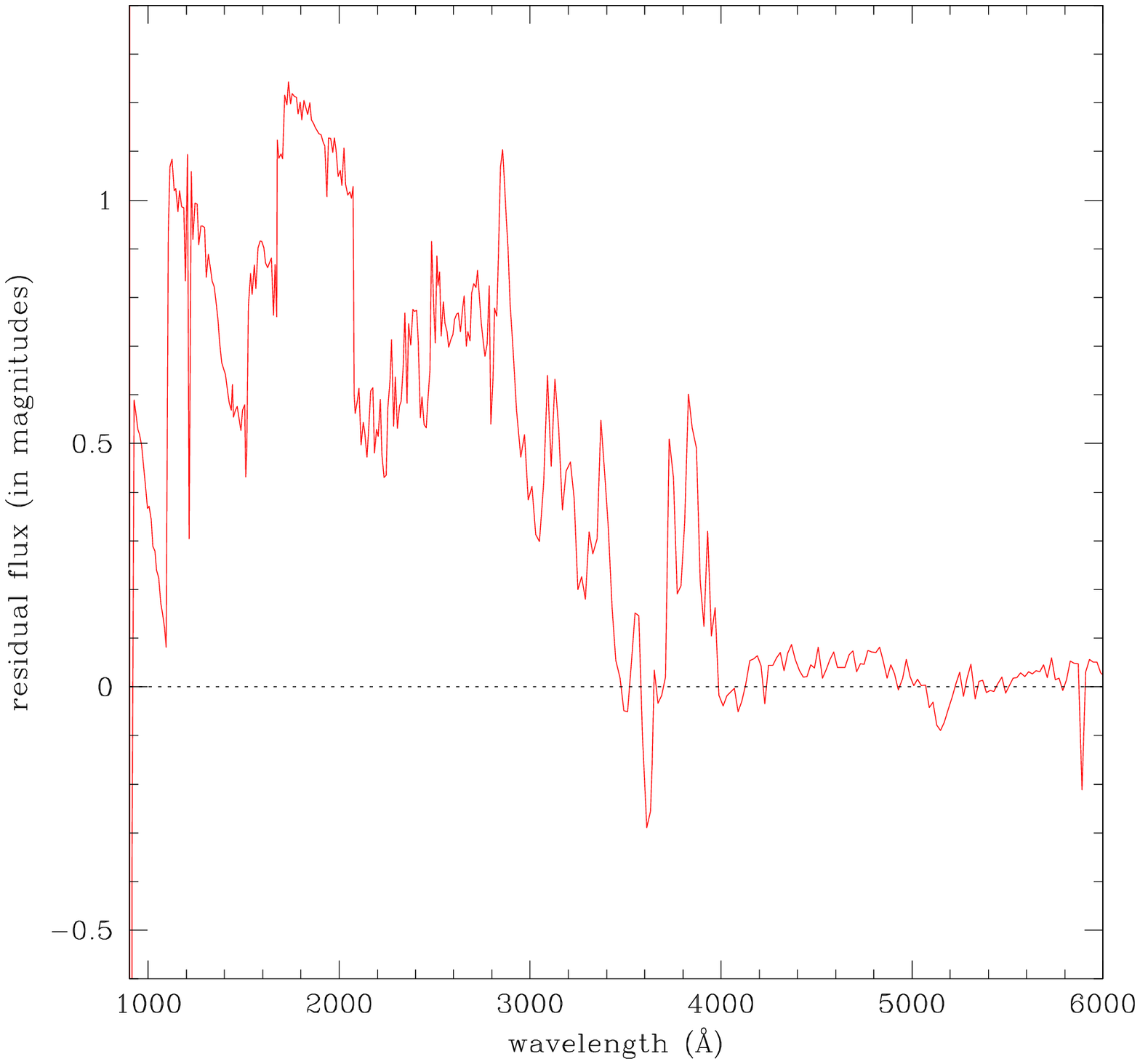}
\end{tabular}
\caption{An example of the age-metallicity degeneracy. In the upper panel we plot the spectral energy distributions (normalized to the $K$ band), of
two simple stellar populations for the ages and chemical compositions as labeled in the figure. We have used the models of \citet{worthey94} for a Salpeter initial mass function. In 
the lower panel we display the flux residual in magnitudes of the SEDs. Note that while the differences are small in the optical interval, in the UV the residuals reach up to one magnitude, indicating that if the age-metallicity degeneracy is present in the UV it is necessarily different from that in the optical}
\label{fig:agemet}
\end{figure}

The optical colors of old populations are affected by the age-metallicity
degeneracy \citep{worthey94}: it implies that the spectrophotometric
properties of an unresolved stellar population can not be distinguished from
those of another population three times older and with half the metal content. 
As an example, we show, in Fig.~\ref{fig:agemet}, the comparison of the
spectral energy distributions (SEDs) [calculated using the models of
\citet{worthey94}\footnote{through the on-line application {\it Dial a
    Galaxy}:\\ http://astro.wsu.edu/worthey/} and normalized to the $K$ band] of two simple stellar 
populations with ages and abundances indicated in the labels of the figure. In the lower planel, we display 
the flux residual in magnitudes of the two energy distributions.

Being arguably amongst the main parameters of a stellar population, there have been a number of studies 
aimed at finding the appropiate features(s) that unambiguously separate the effects of age from those 
of metallicity. In fact, \citet{worthey94} conducted a detailed analysis of optical 
features in the form of spectrocopic indices (the so-called Lick indices) and found that akin the broad band 
colors, the indices, while partially diminishing the AMD, are also degenerate. 

More recently, alternative spectral windows have been proposed as promising tools to lift the AMD. 
In particular, the rest-frame mid-UV flux and colors \citep{yi03,dorman03,kaviraj07} has been investigated 
on the basis that the UV properties are dominated by different stellar types at different evolutionary 
phases (MS) with respect to those dominating the optical (red giants). The overall results indicate that the 
UV indeed helps to better constrain the age of unresolved systems (as would be expected since the MS turn-off 
are much more sensitive to age than the red giant branch), but the determination of chemical composition was 
still better determined by the more sensitive
optical features. The obvious path to further tackle the AMD problem was the use of mid-UV spectroscopic indices, 
however, there was the prevalent concept (somewhat justified, but quantitatively not investigated) that the 
use of synthetic UV indices at the appropiate resolution was still inadequate
for the study of, for example, IUE spectra \citep[see brief discussion in][]{chavez07} and therefore 
investigations of absorption indices was conducted by using, for instance, \citet{kurucz93} low 
resolution grid \citep[e.g.,][]{lotz00}, which appeared more reliable.   

To date, the use of mid-UV synthetic indices is still in its infancy. Suffice
here to mention that the citations to the relevant works where they were
defined \citep{fanelli90} are outnumbered by a factor of 15~(!) by the papers referencing the optical 
indices definition \citep{wortheyetal94}, albeit they were defined roughly at the same time.

\section{Distant Red Galaxies}
Observations at optical and infrared (IR) wavelengths of distant red\footnote{In the context of this paper, we ascribe 
red galaxies to the extremely red objects (EROs), that are intrinsically red, and not to the dust-enshrouded 
star-forming systems.} galaxies (up to redshift of $z \sim 2$) probe the
rest-frame UV range, in particular the 
mid-UV. The first high $z$ red galaxies detected were two faint radio sources from the Lieden-Berkeley Deep 
Survey (LBDS): LBDS 53W091 ($z=$1.55) and LBDS 53W069 ($z=$1.43) \citep{dunlop96,spinrad97,dunlop99}.  
The analysis of these systems was soon a subject of much debate. For instance, \citet{spinrad97} determined 
an age of 3.5~Gyr for LBDS 53W091, which posed complications to explain galaxy
formation under an Einstein-De~Sitter universe. This age was soon contested by a series of authors \citep{bruzual97,heap98,yi00} 
that derived much younger ages ($<$2~Gyr), which allowed for more comfortable estimates for the
formation redshift ($z_{F}$) of the galaxies. Subsequent analyses revived the polemic by confirming the 
first determinations, i.e. ascribing ages in excess of 3~Gyr \citep{nolan03,ferreras04}. Aside from the different 
methodologies used for the age determinations, it was clear that our 
poor knowledge of the UV spectrum of the presumably well understood MS stars \citep[e.g.,][]{peterson01} was 
(and still is to some extent) a major drawback that has prevented the unambiguous determination of the main properties 
(age and chemical composition) of these distant systems. 

More recently, a series of deep surveys have been conducted
\citep{cimatti02,abraham04,mccarthy04} and now include well over 300 systems
with similar spectrophotometric properties as those of the prototypical LBDS 53W091. 
\citet{cimatti08} presented what perhaps is the best spectrum representative of distant red objects. Within the
Galaxy Mass Assembly ultra-deep Sky Survey (GMASS) program, they selected 13 passive galaxies 
(with $1.3 < z < 2.0$) on the basis of their
red UV color, defined as the magnitude difference between two bands (each of
400~\AA\ width) centered at 2900 and 3300~\AA, and constructed a {\it stacked}
spectrum that totalled nearly 500 hours of observing time at the Very Large
Telescope. By comparing that spectrum with single stellar populations (SSPs) from several
population synthesis codes \citep{bruzual03,maraston05}, they determined, from
the rest-frame UV alone, ages that ranged from 0.7 to 2.8~Gyr and metallicities in the range 0.2 to 1.5~$Z_{\odot}$. 
By adding to the comparison near and mid IR photometric data, they significantly constrained the ages to 1--1.6~Gyr and 
found that $Z=Z_{\odot}$ provided the best results. 

\begin{figure*}[!t]
\centering
\includegraphics[scale=0.80]{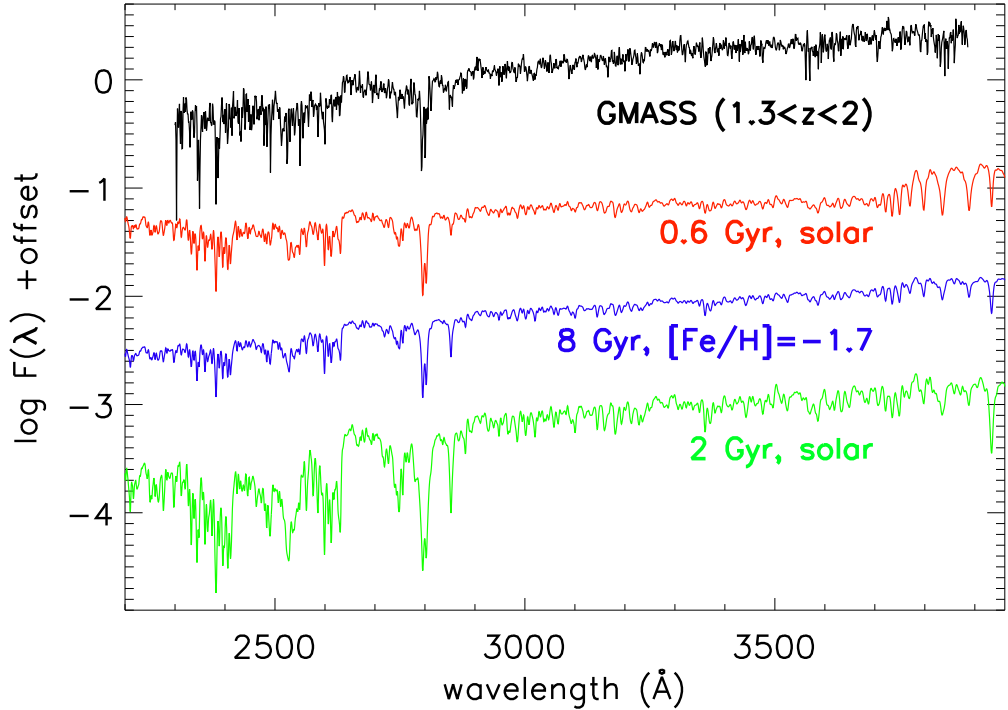}
\caption{GMASS composite spectrum (in black) compared to three theoretical SSP energy distributions 
calculated with {\sc UVBLUE} database and the synthesis code of \citet{bressan94} with the updates described in 
\citet{chavezetal09}. Qualitatively, high $z$ galaxies can be well represented by either a very young (0.6Gyr) and 
solar metallicity or a rather old (8Gyr) subsolar ([M/H]=-1.7) populations}
\label{fig:gmass_seds}
\end{figure*}

In Fig. \ref{fig:gmass_seds}, we show  the GMASS {\it stacked} spectrum of the
13 red galaxies (black) together with three different SSPs of various ages and chemical compositions. As a qualitative 
demonstration of the AMD in the UV, we note that the observed spectrum is very similar to the middle two SSP fluxes 
constructed with quite different parameters.

It is beyond the scope of this paper to discuss any detail on the procedures so far delevoped to establish the 
age and chemical composition of distant systems. We, nevertheless, believe that in general the 
spectrophotometric analysis of distant objects has been carried out with stellar libraries that 
might be inadequate, in particular concerning the spectral resolution and capabilities of 
representing real stars.

\section{Evolved Stellar Populations in the UV}

Back in 2002 the Stellar Atmospheres and Populations Research Group
(GrAPEs--for its designation in spanish) at the Instituto Nacional de
Astrof{\'\i}sica, \'Optica y Electr\'onica initiated a project aimed at providing updated stellar tools for the 
analysis of the UV spectra of a variety of stellar aggregates, mainly evolved ones. The overall project consists in 
four main steps, namely a)- the creation of a theoretical stellar database that we have called {\sc UVBLUE}\footnote{http://www.inaoep.mx/$\sim$modelos/uvblue/uvblue.html}, b)- the comparison of such data base with 
observational stellar data, c)- the calculation of a set of synthetic SEDs of SSPs and their validation through a 
comparison with observations of a sample of Galactic globular clusters, d)- construction of models for dating local 
ellipticals and distant red galaxies.  
In \citet{chavez09}, we presented a summary of the results obtained in steps
(a) and (b) and the reader is referred to that paper and the original
references for a detailed description of the project
\citep{lino05,chavez07}. In what follows, we elaborate on the third step.

\subsection{UV Spectroscopic Indices in Globular Clusters}

In \citet{chavezetal09} we presented the first theoretical analysis of the UV integrated spectra of evolved 
SSPs \citep[see also][for young populations]{maraston09}. 
We focused on particular absorption lines and blends to establish, through the use of spectroscopic indices, their 
behavior in terms of age and chemical composition. We identified several interesting tendencies, such as the low general sensitivity 
of the indices to age and the remarkably distinct behavior of the indices Fe~\textsc{ii}~2332 and Fe~\textsc{ii}~2402,
at super solar regimes (in fact, we propose these indices as a promising tool to establish the age in metal-rich systems).
Synthetic indices were compared to IUE low resolution observations of
prototypical simple populations, i.e. globular clusters, and the results were
highly encouraging, indicating that theoretical SSPs might be confidently used
in the analysis of more complex systems. There were two additional results
that will be important in future analyses: we quantitatively showed that the
presence of hot stars (e.g., blue stragglers and blue horizontal branch (B-HB) objects, which, by the way, are among the 
main contributors to the far-UV rising branch) can significantly dilute the mid-UV absorption indices, and that the 
enhancement of $\alpha$-elements considerably modifies the overall SED of evolved populations.

Based on the results obtained so far, the project at its current stage is now focusing on the detailed analysis of local
(mostly based on IUE observations) and distant evolved systems. This study
will include, in a similar way as \citet{cimatti08}, two steps: we are first
conducting a UV analysis that will be later followed by a panchromatic study using, for instance, the modelling 
machinery developed by \citet{panuzzo05}. We are also carrying out a detailed study of
the far-UV indices and its validation process (as we did in the mid-UV) with the main goal 
of determining the metallicity of the objects responsible of the far-UV up turn.

\subsection{The Sun, M32, and Distant Red Galaxies from a Purely UV Perspective}
In \citet{bertone09}, we presented a preliminary study of the mid-UV spectra
of the Sun and M32 and determined, through a $\chi_\nu^2$ analysis, their age
and chemical composition. Briefly, this analysis consisted in comparing the
observed SEDs of the Sun, extracted from the UARS/SUSIM
archive\footnote{http://daac.gsfc.nasa.gov/data/dataset/UARS/SUSIM}, and that
of M32, taken with the Faint Object Spectrograph onboard the Hubble Space
Telescope (program 1D=6636; PI: M. Gregg), to a set of theoretical integrated
spectra calculated with the synthesis code developed by \citet{buz89}. In the
synthesis code, we have incorporated the {\sc UVBLUE} stellar library and
considered a red HB morphology with a Salpeter initial mass function
($s$=2.35).

The results are listed in Table~\ref{tab:sunm32} \citep[see][for more
details]{bertone09}. Interestingly, we obtained that for the Sun (or,
equivalently, for a population whose mid-UV spectrum is dominated by stars
like the Sun), the absolute chi-square minimum is found for the solar
metallicity and an age of 10.1 Gyr. This result is in remarkably good
agreement with the solar age at the turn-off
\citep[e.g.,][10.5~Gyr]{jorg91}. Similarly, for the central region of M32 we
found a {\em best} fiducial age for the stellar population of the central
region of M32 of 3.64~Gyr, at solar metallicity. This result is, again, in
quite good agreement with the generally accepted age of 3--4~Gyr at solar (or
slightly super-solar) metallicity \citep[see, e.g.,][]{wor04,sch04}.

We have to note, however, that even though we obtained a ``best value", in many
instances, it is difficult to assess the significance of the difference of the
minimum $\chi_\nu^2$ values at the different metallicities. For example, the lowest $\chi_\nu^2$ for Z=0.01, 
solar, and 0.03 for M32 are quite similar. Moreover, these results indicate that the small difference 
in the metal content between Z=0.01 and 0.017 produces a tremendous shift in the age of about 10 Gyr. 
This indicates that the age-metallicity degeneration is clearly present in UV spectra of stellar systems 
and, as mentioned before, operates in a different manner with respect to the optical.

A provocative exercise is to try to determine the age and chemical composition of distant ellipticals 
from a similar analysis, this is, solely based on their mid-UV spectrum. One of course can brandish 
that a panchromatic analysis (UV$+$optical$+$IR) should lead to an unambiguous determination of the parameters. 
Nevertheless, allow us, for now, to assume that we can not complement our UV data with optical and IR (or (sub)-mm data) 
as might be the case for the distant EROs for which we only have the IR 
fluxes (used for their selection from the surveys). Let us also assume that
the UV light is indeed dominated by MS stars at the turn-off, as
would be expected for systems such as globular clusters with red-clumped HBs
or galaxies devoid of field counterparts of B-HB stars and their progeny. In other words, the global 
shape of the SEDs is not modified and the mid-UV spectroscopic 
indices are not diluted by the presence of hotter stars than the MS turn-off. This latter assumption might 
be tested with the measurement of the excess in the far-UV. Let us finally take for granted that the co-added 
spectrum depicted in Fig.~\ref{fig:gmass_seds} is representative of distant single objects.  

\begin{figure}[!t]
\centering
\includegraphics[scale=0.50]{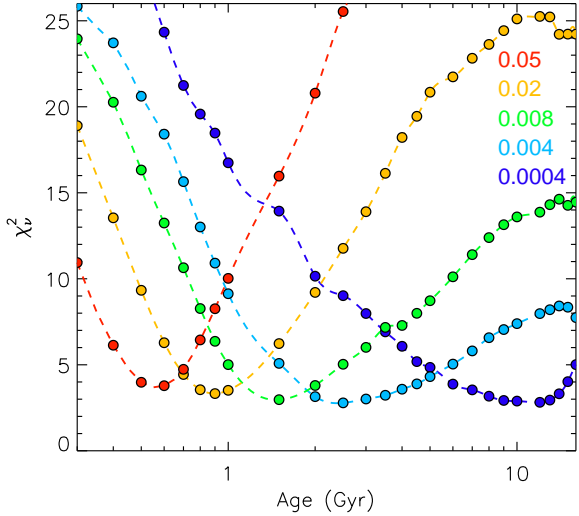}
\caption{Distribution of $\chi_\nu^2$ vs. age at different metallicities for the composite GMASS spectrum 
of \citet{cimatti08}. Different colors stand for five different metallicities, as described in the figure label.}
\label{fig:gmass_fit}
\end{figure}

Figure~\ref{fig:gmass_fit} shows the $\chi_\nu^2$ distribution vs.~age for the GMASS spectrum. For the analysis of this 
spectrum we have used the synthesis code of \citet{bressan94} with the updates described in \citet{chavezetal09}. 
The metallicities considered for this case range from $Z$=0.0004 to 0.05 (as labelled in the figure). The reason 
for using this code is that it includes younger ages ($<$2~Gyr) than that of \citet{buz89}. The analysis
indicates (see results in Table~\ref{tab:gmass}) that the smallest $\chi_\nu^2$ is obtained for an age 
of 2.40~Gyr and a chemical composition of $Z$=0.004. Nevertheless, analogously to the trends found for 
the Sun and M32, the minima are still more uneffective to segregate which of the results is more reliable. 

At present, we are conducting the detailed analysis of the full sample of elliptical galaxies observed by 
IUE and instrisically distant red objects. The aim is not only to test other statistical methods (aside from 
the reduced $\chi_\nu^2$), but to also test the validity of the different assumptions upon which the studies 
can de carried out.

\begin{table}[!t]
\begin{center}
\label{tab:sunm32}
\caption{Age and metallicity for the SUN and M\,32.}{\small
\begin{tabular}{lr@{}lrr@{}lr}
\tableline
\noalign{\smallskip} 
  & \multicolumn{3}{c}{Sun}  &  \multicolumn{3}{c}{M\,32}\\
Z & \multicolumn{2}{c}{Age (Gyr)}  &  $\chi_\nu^2$ & \multicolumn{2}{c}{Age (Gyr)}  &  $\chi_\nu^2$ \\
\noalign{\smallskip}
\tableline
\noalign{\smallskip}
0.0001 &  9.575 & & 4347.0 & 10.000 & & 153.9 \\
0.001  & 15.000 & & 2542.6 & 15.000 & &  84.7 \\
0.010  & 15.000 & & 82.6   & 13.040 & &  19.5 \\
0.017  & 10.115 & $^{+1.620}_{-1.255}$ & 8.9 & 3.640 & $^{+0.725}_{-0.440}$ & 14.3 \\
0.03   &  7.465 & $^{+1.210}_{-0.995}$ & 9.9 & 2.790 & $^{+0.515}_{-0.545}$ & 15.1 \\
0.1    &  6.000 & & 57.6   &  6.000 & & 82.2 \\
\noalign{\smallskip}
\tableline
\end{tabular}
}
\end{center}
\end{table}

\begin{table}
\centering \caption{Best fit parameters for GMASS galaxies}
\label{tab:gmass}       
\begin{tabular}{crr}
\hline \noalign{\smallskip}
 Z       & Age (Gyr)   &  $\chi_\nu^2$        \\
\noalign{\smallskip}\hline\noalign{\smallskip}
0.0004  &  11.75  & 2.81     \\
0.0040  &  2.40   & 2.77     \\
0.0080  &  1.45   & 2.96     \\
0.0200  &  0.90   & 3.32     \\
0.0500  &  0.55   & 3.69     \\
\noalign{\smallskip}
\hline
\end{tabular}
\end{table}

\section{A Wish List for the World Space Observatory-UV}
In the context of what we have discussed above, the WSO-UV \citep{shustov09} will undoubtedly impact our knowledge
on the UV properties of evolved populations. With its large aperture (as compared to its UV predecesors) and enhanced detectors
sensitivities will: 

\begin{itemize}
\item Significantly increase the quality (and quantity) of stars enabling us to construct a 
robust empirical database. As quoted by \citet{chavez07} and \citet{maraston09}, the IUE stellar library has prevailed as the 
most complete for the analyses of stellar aggregates. We, however, badly need to cope with the pausity in the coverage of the 
parameter space, particularly the metallicities.

\item Increase the number of globular clusters (and old open clusters) to re-test the adequacy of SSP models. Of fundamental importance will 
be to empirically judge the effects of non solar-scaled abundances and to better assess the impact of the horizontal branch morphology on mid-UV 
spectroscopic indices. 

\item Cast light on the nature of the objects giving rise to the far-UV up-turn. At this wavelength the IUE-archive includes a rather small
number of elliptical galaxies and none with data of enough quality to firmly establish the chemical composition of the undelying population of 
hot stars in these systems.

\end{itemize}

\acknowledgments
MC is pleased to aknowledge the organizing committee for the invitation to attend the conference 
{\it UV Universe 2010} and for partial financial support. MC and EB would like to thank financial support from
CONACyT through the grant 49231-E.


\end{document}